\documentclass{pr-imfp00}
\usepackage{amssymb,epsfig}

\def\be{\begin{equation}}
\def\ee{\end{equation}}
\def\bea{\begin{eqnarray}}
\def\eea{\end{eqnarray}}

\newcommand{\lsim}{\mathrel{\mathop{\kern 0pt \rlap
  {\raise.2ex\hbox{$<$}}}
  \lower.9ex\hbox{\kern-.190em $\sim$}}}
\newcommand{\gsim}{\mathrel{\mathop{\kern 0pt \rlap
  {\raise.2ex\hbox{$>$}}}
  \lower.9ex\hbox{\kern-.190em $\sim$}}}

\begin{document}

\title{NEUTRINOS }
\author{J. BERNAB\'{E}U}
\address{Jose.Bernabeu@uv.es}
\maketitle

\bigskip

\address{Department de F\'{\i}sica Te\`{o}rica, Universitat de Val\`{e}ncia}


\abstracts{ In these lectures the following topics are considered: historical remarks and general
 properties, Dirac and Majorana neutrino masses, effective lagrangian approach, the
seesaw mechanism, the number of active left-hauded neutrino species, the light neutrino
mass matrix, the direct measurement of neutrino masses, double beta decay, neutrino
oscillations in vacuum and neutrino oscillations in matter. }

\section{Introduction.}

Recently, important events occured in neutrino physics: the Super-Kamiokande
Collaboration reported a strong evidence for neutrino oscillations in their
atmospheric neutrino data $\left[ 1\right] $. The results of the neutrino
experiments will be discussed in the Lectures by Dr. L. Di Lella in these
Proceedings, including future projects. In the present lecture notes a
number of issues pertaining to neutrino physics are considered: general
properties, Dirac and Majorama masses, effective lagrangian approach, the
seesaw mechanism, the number of active left-handed neutrino species, the
light neutrino mass matrix, the direct measurement of neutrino masses,
double beta decay, neutrino oscillations in vacuum and neutrino oscillation
in matter.

Neutrinos play a very important role in various branches of subatomic
physics as well as in astrophysics and cosmology. The neutrino mass problem
is the first window to physics beyond the Standard Model of particle
physics. The smallness of neutrino mass is likely related to the existence
of high mass scales, so high that their direct experimental study is not
accessible. The neutrino mass studies may provide some clues to the general
problem of fermion mass generation.

It was in 1930 when Wolfgang Pauli wrote his letter addressed to te ``Liebe
Radioaktive Damen und Herren'' (Dear Radioactive Ladies and Gentlemen), the
participants of a meeting in T\"{u}bingen. He put forward the hypothesis
that, besides electrons and protons, a new particle exists as ``constituent
of nuclei'', the ``neutron'' $\nu $, able to explain the continuous spectrum
of nuclear beta decay $(A,Z)\rightarrow (A,Z+1)+e^{-}+\nu $. The neutrino is
light (in Pauli's words: ``the mass of the neutron should be of the same
order as the electron mass''), neutral (in today's language, neutral in
electric charge as well as in colour charge) and has spin 1/2. In 1934,
Fermi $\left[ 2\right] $ gave to the neutrino its name and first proposed
the four-fermion theory of beta decay, in terms of charged currents. In 1956
the neutrino was observed for the first time by Cowans and Reines $\left[ 3%
\right] $ in a reactor experiment, by means of its direct detection by its
interaction with matter.

\section{Neutrino mass and helicity}

The first theoretical argument in favour of a vanishing neutrino mass was
provided in 1957 by the two-component neutrino theory $\left[ 4\right] $,
formulated by Landau, Lee and Yang, and Salam. After the discovery of parity
non-conservation by weak interactions, the four-fermion hamiltonian of beta
decay had to be written in terms of currents containing both parity even and
parity odd components.

\bigskip

\begin{equation}
{\cal H}=\sum_{i}(\bar{p}O^{i}n)(\bar{e}O_{i}[G_{i}+G_{i}^{\prime }\gamma
_{5}]\nu )+{\rm h.c.}  \label{eq:num1}
\end{equation}
with the Dirac matrices O$^{i}=1,\gamma ^{\alpha },\sigma ^{\alpha \beta
},\gamma ^{\alpha }\gamma _{5}^{{}},\gamma _{5}^{{}}$. For any Dirac fermion
field $\psi $, one can write $\psi =\psi _{L}+\psi _{R}$, with $\psi _{L,R}=%
\frac{1\mp \gamma _{5}^{{}}}{2}\psi $ the left and right chiral fields. If
the neutrino mass is zero, the chiral projectors $\frac{1\mp \gamma _{5}^{{}}%
}{2}$ become the helicity projection operators and helicity is Lorentz
invariant for zero mass. In the two component neutrino framework, one
assumes that only $\nu _{L}$ enters into $H$. As a consecuence, neutrinos
are produced with negative helicity whereas antineutrinos have positive
helicity. This picture corresponds to maximal violation of charge
conjugation (no left-handed antineutrino) and of parity (no rigth-handed
neutrino). The neutrino helicity was measured in 1957 in a cellebrated
experiment by Goldhaber et al. $\left[ 5\right] $: the electron capture
reaction

\bigskip

\begin{eqnarray}
e^{-}+\phantom{|}^{152}\!Eu\rightarrow \nu +\phantom{|}^{152}\!Sm^{\ast } &&
\label{eq:num2} \\
&\rightarrow &\!\!\!\!\!\!\!^{152}\!Sm+\gamma  \nonumber
\end{eqnarray}
leads to a polarized final state which transmits its polarization to the $%
\gamma $-ray of the subsequent decay. The neutrino is thus left-handed,
implying $G_{i}^{\prime }=-G_{i}$ in Eq. (1). In this view, the neutrino
would be an exceptional particle with $m_{\nu }=0$ and the neutrino field is
$\nu _{L}$.

Soon afterwards the V-A theory of weak interactions was proposed $\left[ 6%
\right] $, in which the left-handed chiral fields of ALL fermions enter into
$H$

\bigskip

\begin{equation}
{\cal H}=\frac{G}{\sqrt{2}}4 (\bar{p}_L \gamma^{\alpha} n_L) (\bar{e}_L
\gamma_{\alpha} \nu_L)+{\rm h.c.}  \label{eq:num3}
\end{equation}

For massive particles, one should carefuly distinguish between chirality and
helicity. Let us consider the weak decay $\pi ^{+}\rightarrow l^{+}+\nu _{l}$
of pions at rest, with $l=\mu ,e$ and $\nu _{l}$ the corresponding neutrino.
This semileptonic decay is again described by the product of an hadronic
current and a leptonic current. In the V-A structure, the emitted $\nu _{l}$
is left-handed. For $m_{\nu }=0$, this also means negative helicity.
Conservation of total angular momentum requires $l^{+}$ to be of negative
helicity too. However, the $l^{+}$ are antiparticles and the V-A theory
predicts that they are produced with right-handed chirality. As a
consequence, the probability amplitude of this process is proportional to
the admixture of negative helicity of the charged lepton in its right-handed
chirality, i.e., to its mass $m_l$. Including the phase space factor, one
thus expects

\bigskip

\begin{equation}
R_{\pi }\equiv \frac{\Gamma (\pi ^{+}\rightarrow e^{+}+\nu _{e})}{\Gamma
(\pi ^{+}\rightarrow \mu ^{+}+\nu _{\mu })}=\left( \frac{m_{e}}{m_{\mu }}%
\right) ^{2}\left( \frac{m_{\pi }^{2}-m_{e}^{2}}{m_{\pi }^{2}-m_{\mu }^{2}}%
\right) ^{2}=1.28\times 10^{-4}.  \label{eq:num4}
\end{equation}
When the 4\% effect of radiative corrections is considered, this result is
in agreement with the experimental value $\left[ 7\right] $ $R_{\pi
}=(1.230\pm 0.004)\times 10^{-4}$. This manifestation of the chirality
suppression follows from the V-A character of the weak current, with a
neutrino mass much smaller than that of the charged lepton.

The argument of the two component neutrino theory disappears once it is
realized that the left-handed chiral fields are the ones which participate
in weak interactions for all fermions. In this sense there is nothing
special for neutrinos. There is, however, a crucial difference: except for
neutrinos, the right handed chiral fields have to exist in Nature for
charged leptons (QED) and for quarks (QED +QCD). Once both left-and
right-handed chiral fields are present in particle theory, one has a Dirac
mass term for up and down quarks and for (down) charged leptons. In the
Standard Model, in which the origin of mass comes from spontaneous symmetry
breaking, the Yukawa interaction $\left[ 8\right] $ between the left-handed
doublet $\psi _{L}$ and the right-handed singlet $l_{R}$ is (in matrix
notation in fermion family space)

\bigskip

\begin{equation}
{\cal L}_{Y}^{(\ell )}=-\frac{\sqrt{2}}{v}\bar{\psi}_{L}M^{(\ell )}\ell
_{R}\varphi +{\rm h.c.}  \label{eq:num5}
\end{equation}
with $\varphi $ the scalar doublet. After spontaneous symmetry breaking,

\bigskip

\begin{equation}
\varphi \stackrel{SSB}{\rightarrow }\left\{
\begin{array}{c}
0 \\
\frac{v+H}{\sqrt{2}}
\end{array}
\right\} \Rightarrow {\cal L}_{{\rm mass}}^{(\ell )}=-\bar{\ell}_{L}M^{(\ell
)}\ell _{R}+{\rm h.c.}  \label{eq:num6}
\end{equation}
with $M^{(l)}$ the Dirac mass matrix for charged leptons.

\section{Dirac versus Majorana neutrinos}

Neutrinos, contrary to other fermions, do not participate in parity
conserving vector-like interactions QED or QCD, and only $\nu _{L}$ is
relevant for weak interactions. There is no need of introducing $\nu _{R}$
into the theory as an independent field. If one does it, against the
``choice'' of the Standard Model, the $\nu _{R}$ is sterile against gauge
interactions and only suffers the Yukawa interaction

\bigskip

\begin{equation}
{\cal L}_{Y}^{(\nu )}=-\frac{\sqrt{2}}{v}\bar{\psi}_{L}M^{(\nu )}\nu _{R}%
\widetilde{\varphi }+{\rm h.c.}  \label{eq:num7}
\end{equation}
where $\widetilde{\varphi }$ is the charge-conjugated of the scalar field.
The forms (5) and (7) are dictated by $SU(2)\times U(1)$ gauge invariance.

Under spontaneous symmetry-breaking,

\bigskip

\begin{equation}
\widetilde{\varphi }=i\tau _{2}\varphi ^{\ast }\stackrel{SSB}{\rightarrow }%
\left\{
\begin{array}{c}
\frac{v+H}{\sqrt{2}} \\
0
\end{array}
\right\} \Rightarrow {\cal L}_{{\rm mass}}^{(\nu )}=-\bar{\nu}_{L}M^{(\nu
)}\nu _{R}+{\rm h.c.}  \label{eq:num8}
\end{equation}
with $M^{(\nu )}$ a Dirac mass matrix for neutrinos. In this alternative,
the neutrinos would be Dirac particles, in total analogy with quarks. The
leptonic sector would have the analogous to the CKM matrix: the MNS matrix$%
\left[ 9\right] $\ and Eq. (8) induces masses and mixings. The mixing is
relevant for charged current interactions, because $M^{(\nu )}\neq M^{(l)}$.
The mixing is, however, irrelevant for the neutral current interaction,
leading to GIM-suppressed flavour changing neutral currents. Besides Eq.
(7), the $\nu _{R}$'s do not appear elsewhere. The matrix $M^{(\nu )}$ leads
to lepton flavour violation, but the Lagrangian is still invariant under a
Global $U$(1)-Gauge Transformation of Total Lepton Number. Neutrinoless
Double Beta Decay would be thus rigorously forbidden in Nature.

As $\nu _{R}$ is sterile for gauge interactions, we can contemplate a second
alternative for neutrinos (forbidden for the other fermions): $\nu _{R}$
does not exist as an independent field. We can ask ourselves: Is it
possible, with only $\nu _{L}$, to generate a non-vanishing neutrino mass? A
priori, the answer to this question is positive, thanks to Majorana$\left[ 10%
\right] $. For the chiral $\nu _{L}$ field, contrary to a Dirac field, its
charge conjugated $\nu _{L}^{c}$ is right-handed, so that one can write a
Majorana mass term

\begin{equation}
{\cal L}_{{\rm mass}}^{({\rm Maj)}}=-\frac{1}{2}\bar{\nu}_{L}M\nu _{L}^{c}+%
{\rm h.c.}  \label{eq:num9}
\end{equation}
with only $\nu _{L}$. For neutrinos, Eq. (9) is not only Lorentz-invariant,
but also $SU$(3)$_{colour}\otimes U$(1)$_{e.m.}$ invariant. It is thus
perfectly legal, contrary to all other fermions carrying conserved charges.
The requirement of anticommutation for the quantum fermion fields leads to
the symmetry condition $M^{T}=M$ for the Majorana mass matrix. $M$ is, in
general, a complex symmetric matrix and it can be diagonalized by means of a
unitary matrix $U$ according to

\bigskip

\begin{equation}
M=UmU^{T}  \label{eq:num10}
\end{equation}
with $m$ the diagonal matrix of mass eigenvalues. Eq. (9) can be written in
terms of the fields $\chi $ with definite mass

\bigskip

\begin{equation}
\left.
\begin{array}{l}
{\cal L}_{{\rm mass}}^{({\rm Maj)}}=-\frac{1}{2}\bar{\chi} m \chi \\
\chi \equiv U^+ \nu_L + (U^+ \nu_L)^c
\end{array}
\right \}  \label{eq:num11}
\end{equation}

As seen, the Majorama field $\chi $ is self-conjugated, satisfying

\bigskip

\begin{equation}
\chi=\chi^c \equiv {\cal C} \bar{\chi}^T .  \label{eq:num12}
\end{equation}

In this alternative, the physical neutrinos of definite mass would be true
neutral particles, with no conserved global lepton number. If a lepton
number $L$ is introduced, Eq.(9) transports two units of this lepton charge:
$\Delta L=2$. Neutrinoless Double Beta Decay would be allowed.

In general, the fermion field $\psi (x)$ can be (Fourier) transformed to
momentum space by means of the spinor $u_{\lambda }(p)$ and its charge
conjugated $C\left[ \overline{u}_{\lambda }(p)\right] ^{T}$

\bigskip

\begin{equation}
\psi (x)=\int \frac{d^{3}p}{(2\pi )^{3/2}}\frac{1}{\sqrt{2p^{0}}}%
\sum_{\lambda }\left\{ c_{\lambda }(p)u_{\lambda }(p)e^{-ipx}+d_{\lambda
}^{+}(p)C[\bar{u}_{\lambda }(p)]^{T}e^{ipx}\right\} .  \label{eq:num13}
\end{equation}
where $c_{\lambda }$ is the particle annihilation operator and $d_{\lambda
}^{+}(p)$ is the antiparticle creation operator. With the decomposition
(13), one can construct the Dirac propagator. The vacuum expectation value
of the time ordered product of the field times its adjoint

\bigskip

\begin{equation}
\langle 0|T\{\psi (x)\overline{\psi }(0)\}|0\rangle  \label{eq:num14}
\end{equation}
is non-vanishing for both Dirac and Majorana neutrinos. It describes
neutrino $\rightarrow $ neutrino propagation.

Is there a possibility of Majorana propagation? This is described by the
vacuum expectation value of the time ordered product of the field times
itself. A non-vanishing value $\left[ 11\right] $ of

\bigskip

\begin{equation}
\langle 0|T\{\psi (x)\psi (0)^{T}\}|0\rangle  \label{eq:num15}
\end{equation}
is only possible IFF $\psi =\chi =C$ $\overline{\chi }^{T}$, i.e., for a
Majorona field. The Majorana condition implies $c_{\lambda }(p)=d_{\lambda
}(p)$, which identifies particle and antiparticle. The Majorana propagator
(15) describes ``neutrino$\rightarrow $antineutrino'' propagation, a
manifestation of the $\Delta L=2$ character of the Majorana mass term.

In the standard model, the Majorana mass term of Eq.(9) cannot be generated
by spontaneous symmetry breaking of a Yukawa interaction with a (doublet)
scalar field. The reason is apparent: Eq. (9) is a triplet in weak isospin,
so one would need an isotriplet scalar field, which does not exist in the
standard model.

As a consequence the standard model, with its particle content and
renormalizable dimension-four operators only, predicts that neutrinos are
massless.

\section{Effective Lagrangian Approach}

In the last 30 years one has seen a deep evolution in the understanding of
quantum field theories, so that the requirement of renormalizability is
taken today with a less dogmatic philosophy. Take the particle content of
the standard model and ask what is the lowest dimension non-renormalizable
operator which still keeps the $SU(2)\times U(1)$ gauge invariance. The
answer to this question is unique, given by the dimension-five operator $%
\left[ 12\right] $ in the leptonic sector

\bigskip

\begin{equation}
{\cal L}_{{\rm eff}}=-\frac{1}{2\Lambda }(\bar{\tilde{\psi}}_{L}\varphi
)F(\psi \widetilde{\varphi }^{+}\psi _{L}),  \label{eq:num16}
\end{equation}
where $\widetilde{\psi }=i\tau _{2}\psi ^{c}=i\tau _{2}C\overline{\psi }^{T}$%
. We can say properly that (16) represents the first window to physics
beyond the standard model. The symmetric $F=F^{T}$ matrix induces mixing in
flavour-space. The coupling $\frac{1}{\Lambda }$ is reminiscent of the scale
of new physics at higher energies. Eq. (16) generates, in addition to lepton
flavour violation, $\Delta L=2$ transitions.

After spontaneous symmetry breaking, Eq. (16) leads to a Majorana mass term
(9) for neutrinos, with the mass matrix

\bigskip

\begin{equation}
M=\frac{v^2}{\Lambda} F .  \label{eq:num17}
\end{equation}

We conclude that this new physics mechanism leads to massive Majorana
neutrinos. Eq. (17) provides a very simple and attractive explanation of the
smallness of neutrino mass. It relates the smallness of $m_{\nu}$with the
existence of a very large mass scale $\Lambda $ compared with the
electroweak scale, given by the vacuum expectation value $v=174$ $GeV$. The
effective Lagrangian approach cannot unveil the origin of $\Lambda $,
because the shorter-distance physics has been integrated out. We will
discuss below the see-saw mechanism based on the introduction of very heavy
right-handed $\nu _{R}$. An alternative to it is suggested by the
Fierz-reordered form of the effective lagrangian(16)

\bigskip

\begin{equation}
{\cal L}_{{\rm eff}}=-\frac{1}{4\Lambda }(\bar{\tilde{\psi}}_{L}F\vec{\tau}%
\psi _{L})(\widetilde{\varphi }^{+}\vec{\tau}\varphi ).  \label{eq:num18}
\end{equation}

The last bracket has the same transformation property of a scalar triplet,
so that a heavy Higgs triplet will do the job as well. This theory is not
very attractive nowadays, and it runs into difficulties with the invisible
Z-width.

Several models of neutrino mass and mixing based on the minimalist approach
of this Section are discussed in Ref. $\left[ 13\right] $.

\section{Seesaw mechanism}

The seesaw mechanism $\left[ 14\right] $ is most natural in the framework of
grand unified theories, such as $SO(10)$, or left-right symmetric models, in
which the rigth-handed $\nu _{R}$ acquires a large Majorana mass $(\Lambda )$
as part of the symmetry breaking scenarios. But it also operates in the
standard $SU(2)\times U(1)$ gauge invariant model, extended to include a
heavy $\nu _{R}$.

The most general mass term includes not only the left-right Dirac mass,
generated by the standard spontaneous symmetry breaking in the Yukawa
coupling, but also a Majorana mass for right-handed neutrinos. As $\nu _{R}$
is sterile (electroweak singlet), this last term introduced by hand keeps
the $SU(2)\times U(1)$ gauge invariance. We can write

\bigskip

\begin{equation}
{\cal L}_{{\rm mass}}^{D-M}=-\bar{\nu}_{R}m_{D}\nu _{L}-\frac{1}{2}\bar{\nu}%
_{R}m_{R}(\nu _{R})^{c}+{\rm h.c.}=-\frac{1}{2}(\bar{n}_{L})^{c}Mn_{L}+{\rm %
h.c.},  \label{eq:num19}
\end{equation}
where the last equality follows from organizing the left handed fields
(twice the number of families) as

\bigskip

\begin{equation}
n_L \equiv \left \{
\begin{array}{c}
\nu_L \\
(\nu_R)^c
\end{array}
\right \}, \quad M= \left (
\begin{array}{cc}
0 & m_D \\
m_D & m_R
\end{array}
\right ).  \label{eq:num20}
\end{equation}

Let us discuss the limit $m_{D}<<m_{R}$ for the one-family case, where $M$
is a $2\times 2$ real symmetric matrix. The diagonalization of $M$ leads to
two Majorana neutrinos with masses

\bigskip

\begin{equation}
m_{1}\approx \frac{m_{D}^{2}}{m_{R}},\quad m_{2}\approx m_{R},
\label{eq:num21}
\end{equation}
and definite CP eigenvalues $\eta _{1}=-1$, $\eta _{2}=1$. The mixing angle
is hierarchical $\theta \simeq m_{D}/m_{R}$. Neglecting this small admixture
between ``active'' and ``sterile'' neutrinos, the Majorana Fields with
definite mass are

\bigskip

\begin{equation}
\nu _{1}\approx i\nu _{L}-i(\nu _{L})^{c};\quad \nu _{2}=\nu _{R}+(\nu
_{R})^{c},  \label{eq:num22}
\end{equation}
corresponding to a light neutrino with $m_{1}<<m_D$ and a heavy neutrino
with $m_{2}>>m_{D}$. This solution is a realization of the discussion of the
previous Section, with the high mass scales $\Lambda $ represented by $m_{R}$%
.

The mass Lagrangian of Eq.(19) violates global lepton number $L$ only by the
right-handed Majorana term -$\frac{1}{2}\overline{\nu }_{R}m_{R}(\nu
_{R})^{c}$, characterized by the large mass. One thus connects the smallnes
of the light neutrino mass to lepton number violation at the high mass scale.

The results for one family can be generalized, so that $m_D$ and $m_{R}$ are
matrices of dimension the number of families. Block-diagonalization of $M$
gives

\bigskip

\begin{equation}
m_{L}\approx -m_{D}m_{R}^{-1}m_{D}^{T};\quad m_{R},  \label{eq:num23}
\end{equation}
for the active and sterile neutrinos, respectively. The subsequent
diagonalization of $m_{L}$ leads to active Majorana neutrinos with an
expected hierarchy of masses $m_{1}<<m_{2}<<m_{3}$, when taking into account
that $m_{D}$ is of the order of quark or lepton masses.

\section{How many active neutrinos?}

The counting of light active left-handed neutrinos is based on the family
structure of the standard model and assuming its predictions of a universal
diagonal neutral current coupling. We have the vertex of Fig.1

\begin{figure}[h!]
\centerline{
\epsfxsize=6cm
\epsffile{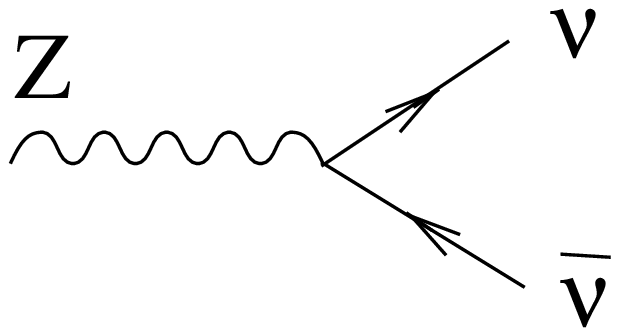}
}
\caption{}
\label{fig1}
\end{figure}

\begin{equation}
j_{Z}^{\mu }=\sum_{\alpha }\bar{\nu}_{\alpha L}\gamma ^{\mu }\nu _{\alpha L}.
\label{eq:num24}
\end{equation}

With this current and the condition $m_{\nu _{i}}<m_{Z}/2$ for the physical
neutrinos, one can use the total width $\Gamma _{Z}$of the $Z$-boson to
extract $N_{\nu }$. A ``nearly'' model independent method can be put forward
as follows

\bigskip

\begin{equation}
N_{\nu }=\frac{\Gamma _{{\rm inv}}}{\Gamma _{\nu }}\equiv \frac{1}{\Gamma
_{\nu }}(\Gamma _{Z}-\Gamma _{h}-3\Gamma _{\ell })=\frac{\Gamma _{\ell }}{%
\Gamma _{\nu }}\left[ \sqrt{\frac{12\pi R_{h\ell }}{\sigma _{h}^{0}m_{Z}^{2}}%
}-R_{h\ell }-3\right] ,  \label{eq:num25}
\end{equation}
where $\Gamma _{inv}$is the invisible width, $\Gamma _{h}$ the total
hadronic width and $\Gamma _{l}$ one of the charged lepton widths. The last
equality leads to a bracket with only experimental inputs: the hadronic
cross section $\sigma _{h}^{0}$ at the peak of mass $m_{Z}$ and the hadronic
to leptonic ratio $R_{hl}$ of widths. The factor in front has to be taken
from the standard model, but the ratio $\Gamma _{l}/\Gamma _{\nu }$ is free
from big universal radiative corrections. The precise maping of the $Z$ line
shape from the four experiments of LEP1 facility shows a clear demonstration
of $N_{\nu }=3$. Using the strategy of Eq.(25), one obtains $\left[ 7\right]
$

\begin{equation}
N_{\nu}=2.994\pm 0.012 .  \label{eq:num26}
\end{equation}

\section{The light neutrino mass matrix}

In the three-flavour framework, without light sterile neutrinos, the
relevant mixing matrix is that indicated by $m_{L}$ in Eq.(23), whatever its
origin is. In a hierarchical solution for the three neutrino masses, $\Delta
m_{32}^{2}\simeq \Delta m_{31}^{2}$could be identified with the mass
difference relevant to neutrino oscillations in the atmospheric neutrino
data, whereas $\Delta m_{21}^{2}<<\Delta m_{32}^{2}$ would be associated
with the solar neutrino problem. The two-flavour analysis is, in fact, a
good first approximation to the results of the three-flavour studies $\left[
15\right] $ because the mixing angle $\theta _{13}$ is constrained to be
rather small by the CHOOZ data $\left[ 16\right] $ $\left| U_{e3}\right|
^{2}<0.02$.

With the assumptions of this non-participation of $\nu _{e}$ in atmospheric
oscillations plus a maximal atmospheric neutrino mixing, compatible with the
experiment $\left[ 1\right] $, we can construct the real and orthogonal
diagonalizing matrix $U$ as

\bigskip

\begin{eqnarray}
U&=&\left[
\begin{array}{ccc}
1 & 0 & 0 \\
0 & 1/\sqrt{2} & -1/\sqrt{2} \\
0 & 1/\sqrt{2} & 1/\sqrt{2}
\end{array}
\right ] \left[
\begin{array}{ccc}
c & -s & 0 \\
s & c & 0 \\
0 & 0 & 1
\end{array}
\right ]  \nonumber \\
&=& \left[
\begin{array}{ccc}
c & -s & 0 \\
s/\sqrt{2} & c/\sqrt{2} & -1/\sqrt{2} \\
s/\sqrt{2} & c/\sqrt{2} & 1/\sqrt{2}
\end{array}
\right ]  \label{eq:num27}
\end{eqnarray}

As a result of $U_{e3}=0$, the theory is CP-symmetric and $U$ can be chosen
real. It has been constructed as a rotation in the (12)-plane and then a
rotation by $\pi /4$ in the (23)-plane. The factorization of atmospheric and
solar mixings will be lost once $U_{e3}\neq 0$ is allowed. This in turn
induces CP violation in the leptonic sector, through a phase that could not
be rotated away. The determination of these parameters is the objective of
the long term neutrino projects.

Solar neutrino experiments still allow several solutions for the rotation in
the (12)-plane. Once determined, this information plus the neutrino spectrum
provide an empirical mass matrix from

\[
m_{L}=U\left(
\begin{array}{ccc}
m_{1} &  &  \\
& m_{2} &  \\
&  & m_{3}
\end{array}
\right) U^{T}
\]

\section{Direct measurement of neutrino mass}

Neutrino oscillations (see below) constitute the most precise method to
search and measure neutrino mass differences. In order to determine the
absolute mass scale, one needs other observables. Fermi proposed $\left[ 2%
\right] $ a kinematic search of neutrino mass from the hard part of the beta
spectra in $^{3}H$ beta decay. With some abuse of language, this search has
produced an upper limit to the electron neutrino mass. The electron neutrino
is a weak state. Due to mixing, it has no definite mass. For the mixing of
Eq.(27), the discussion of this section applies to $\nu _{1}$ with
probability $c^{2}$ and to $\nu _{2}$ with probability $s^{2}$.

For ``allowed'' nuclear trnsitions, the nuclear matrix elements do not
generate any energy dependence, so that the electron spectrum is given by
phase space alone

\bigskip

\begin{equation}
\frac{dN}{dT}=CpE(Q-T)\sqrt{(Q-T)^{2}-m_{\nu }^{2}}F(E),  \label{eq:num28}
\end{equation}
where $E=T+m_{e}$, $Q$ is the maximum energy and $F(E)$ the Fermi function
which incorporates final state Coulomb interactions.

The ``classical'' decay $^{3}H\rightarrow ^{3}He+e^{-}+\overline{\nu }_{e}$
is a superallowed transition with a very small energy release $Q=18.6$ $KeV$%
. In the Kurie plot

\bigskip

\begin{equation}
K(T)\equiv \sqrt{\frac{dN}{dT}\frac{1}{pEF(E)}}\propto \sqrt{(Q-T)\sqrt{%
(Q-T)^{2}-m_{\nu }^{2}}},  \label{eq:num29}
\end{equation}
a non-vanishing neutrino mass $m_{\nu }$ provokes a distorsion from the
straight-line T-dependence at the end point of the spectrum, in such a way
that $m_{\nu }=0\rightarrow T_{\max }=Q$ whereas $m_{\nu }\neq 0\rightarrow
T_{\max }=Q-m_{\nu }$. This is shown in Fig.2

\begin{figure}[h!]
\centerline{
\epsfxsize=8cm
\epsffile{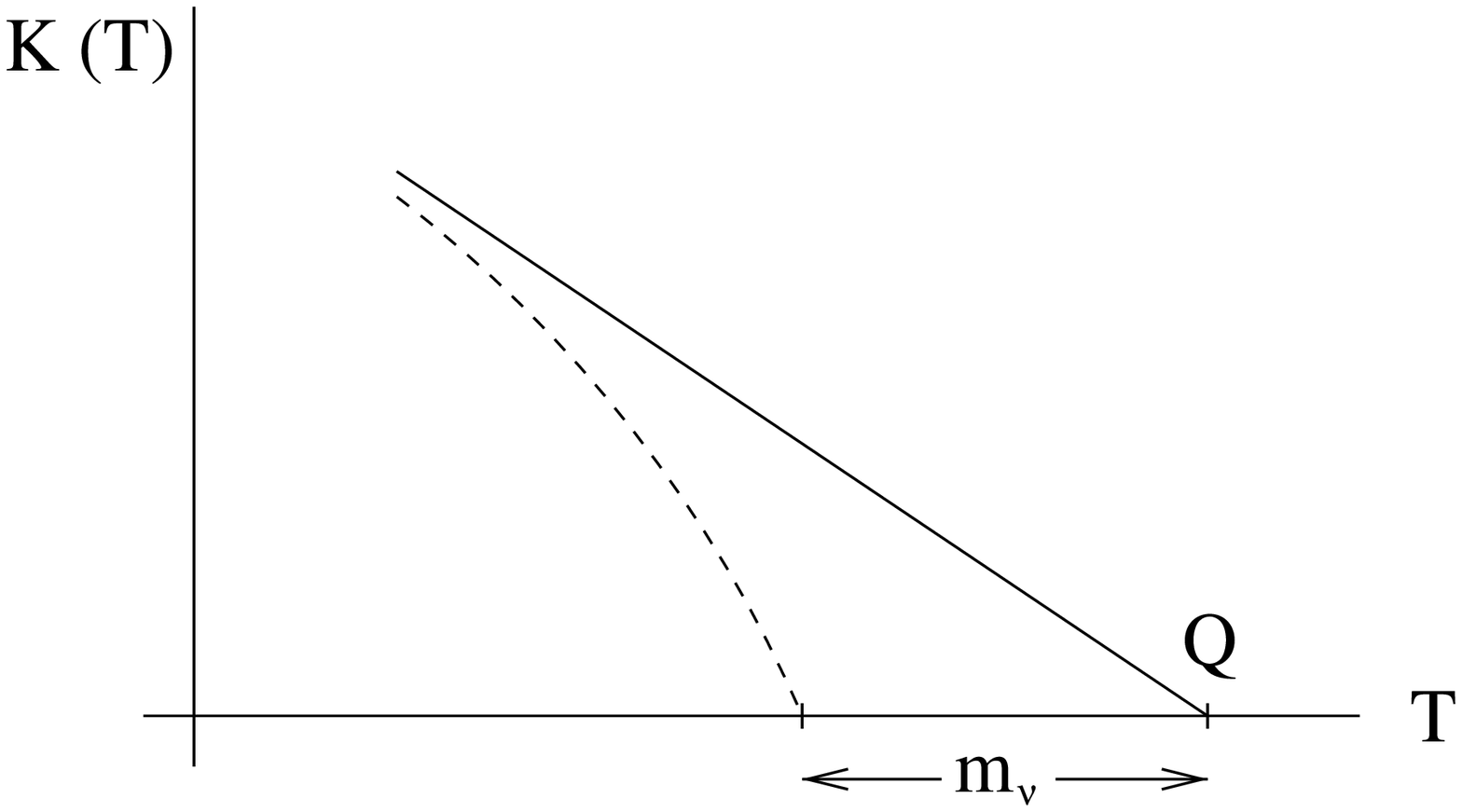}
}
\caption{}
\label{fig2}
\end{figure}

The experimental spectrum is fitted by $m_{\nu }^{2}$ and many other
parameters ($Q$, background term, normalization, ...). The most precise
Troitsk and Mainz experiments $\left[ 7\right] $ give no indication in
favour of $m_{\nu }\neq 0$. One has the upper limit $m_{\nu }<2.5$ $eV$ at
95\% c.l.

\section{Neutrinoless Double Beta Decay}

One of the main pending questions in neutrino physics is: Are neutrinos
Dirac or Majorana particles? Or, equivalently: Is the neutrino its own
antiparticle? The identity of $\nu $ and $\overline{\nu }$ would mean that
both $\nu $ and $\overline{\nu }$ interact with matter in the same way. Do
they? A simple inspection to the experimental cross section $\left[ 7\right]
$ in the region of energies in which the behaviour is linear with the energy
in the lab shows that there is a difference: the neutrino cross section is
about twice the result for antineutrinos. This comparison is, however, not
relevant for our question, because the difference in the total cross section
for neutrinos and antineutrinos is due merely to the different polarizations
of the beams. The so-called neutrinos are left-handed whereas the so-called
antineutrinos are right-handed and the cross section is helicity dependent.

Contrary to the situation described in the total cross section, a test of
the Majorana condition $c_{\lambda }(p)=d_{\lambda }(p)$ (in Eq.(13 ) )
needs the preparation of neutrinos and antineutrinos of the same helicity.

The way to study this problem is based on the $\Delta L=2$ transition
implied by the Majorana mass term. The neutrinoless double beta decay process

\begin{equation}
(A,\,Z)\rightarrow (A,\,Z+2)+e^{-}+e^{-}  \label{eq:num30}
\end{equation}
was proposed by Furry in 1939 $\left[ 15\right] $ and becomes allowed for
Majorana neutrino virtual propagation. It is described by the diagram of
Fig.3

\begin{figure}[h]
\centerline{
\epsfxsize=8cm
\epsffile{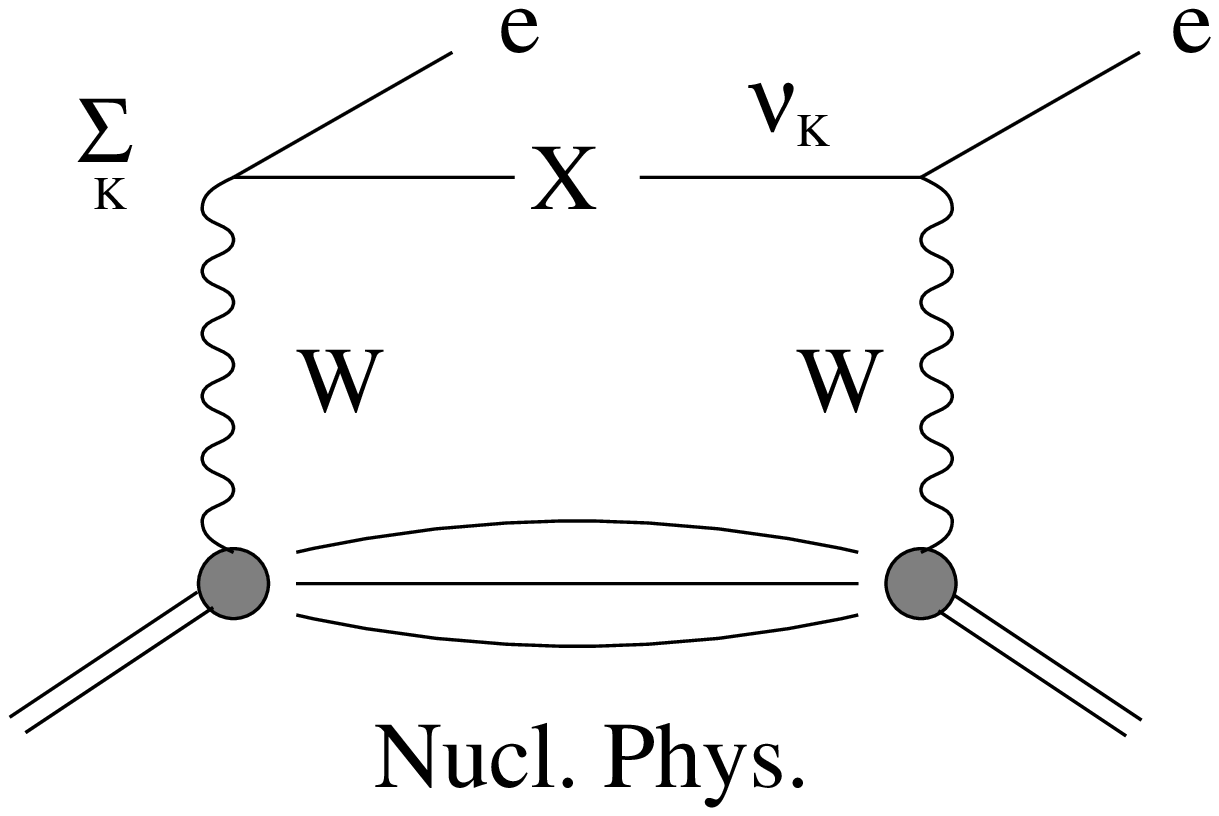}
}
\caption{{}}
\label{fig3}
\end{figure}
as a second order weak interaction amplitude. It becomes a source of nuclear
instability for selected even-even nuclei in which the single beta decay is
energetically forbidden. I show the level diagram corresponding to the decay
of $^{76}Ge$ in Fig.4

\begin{figure}[h!]
\centerline{
\epsfxsize=8cm
\epsffile{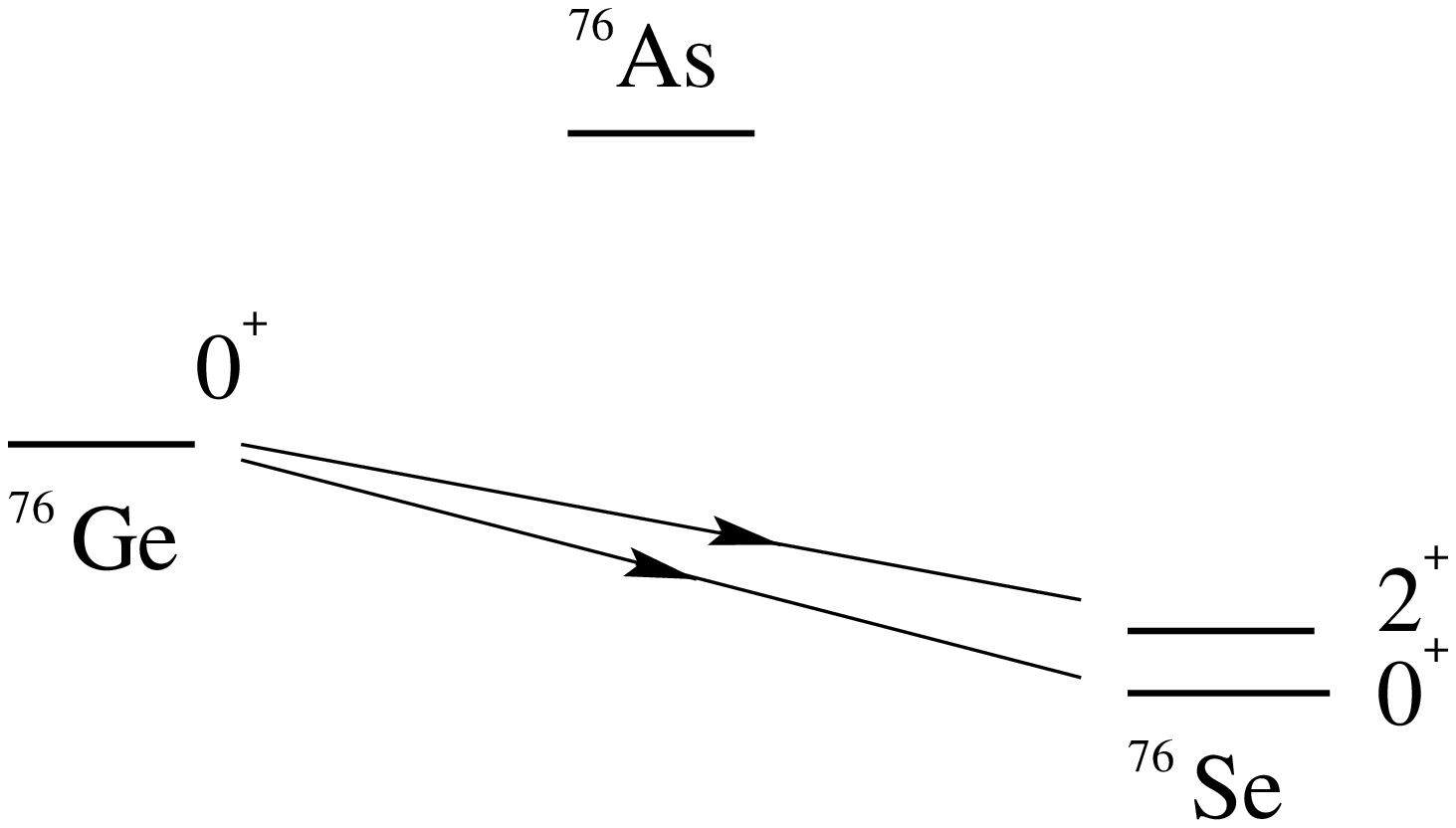}
}
\caption{}
\label{fig4}
\end{figure}

The neutrino propagator is here

\begin{eqnarray}
\widehat{\nu _{eL}(x_{1})\nu }_{eL}^{T}(x_{2}) &=&-\sum_{k}U_{ek}^{2}\frac{%
1-\gamma _{5}}{2}\widehat{\chi _{k}(x_{1})\bar{\chi}}_{k}(x_{2})\frac{%
1-\gamma _{5}}{2}C  \nonumber \\
&=&\sum_{k}U_{ek}^{2}m_{k}\frac{-i}{(2\pi )^{4}}\int d^{4}p\frac{%
e^{ip(x_{1}-x_{2})}}{p^{2}-m_{k}^{2}}\frac{1-\gamma _{5}}{2}C.
\label{eq:num31}
\end{eqnarray}

If $m_k$ are small, compared to the momenta relevant for nuclear physics
excitations, the neutrino masses can be neglected in the denominator of the
propagator. The amplitude of the process is then factorized in its different
ingredients

\begin{equation}
{\rm Amp}[\beta\beta_{0\nu}]=\langle m_{\nu}\rangle({\rm Phase \ Space}) (%
{\rm Nuclear \ Physics}).  \label{eq:num32}
\end{equation}

The quantity of primary interest in neutrino physics is the average neutrino
mass $<m_{\nu }>$

\begin{equation}
\langle m_{\nu}\rangle=\sum_{k} U_{e k}^2 m_k .  \label{eq:num33}
\end{equation}

This result shows that the main ingredient to produce an allowed $(\beta
\beta )_{0\nu }$ is the massive Majorana neutrino character. Even without
mixing, the process is still allowed. In presence of mixing, there are
contributions of the different physical neutrinos to $<m_{\nu }>$,
contributions which can cancel each other. Even with CP-conserving
interactions, the contributions of different CP eigenvalues $\eta _{k}$
appear as

\begin{equation}
\langle m_{\nu}\rangle=\sum_{k} |U_{e k}|^2 m_k \eta_k .  \label{eq:num34}
\end{equation}

Besides these properties, the result (32) shows the dependence of the
amplitude with the absolute neutrino masses, not with the mass differences.
Under favourable circunstances, a positive signal of the $(\beta \beta
)_{0\nu }$ process could be combined with results of neutrino oscillation
studies to determine $\left[ 18\right] $ the absolute scale of neutrino
masses.

The most stringent experimental limit on $<m>$ at present is obtained by the
Heidelberg-Moscow collaboration $\left[ 19\right] $, $<m><0.35eV$, running
in the Gran Sasso underground laboratory. There are prospects to reach
sensitivities of $10^{-2}eV$ in the near future.

To compare with, I should add that the two-neutrino double beta decay

\begin{equation}
(A,\,Z)\rightarrow (A,\,Z+2)+(2e^{-})+(2\bar{\nu}_{e})  \label{eq:num35}
\end{equation}
is allowed, although rare, even if global lepton number is conserved in
second order weak interaction decays as shown in Fig.5

\begin{figure}[h!]
\centerline{
\epsfxsize=8cm
\epsffile{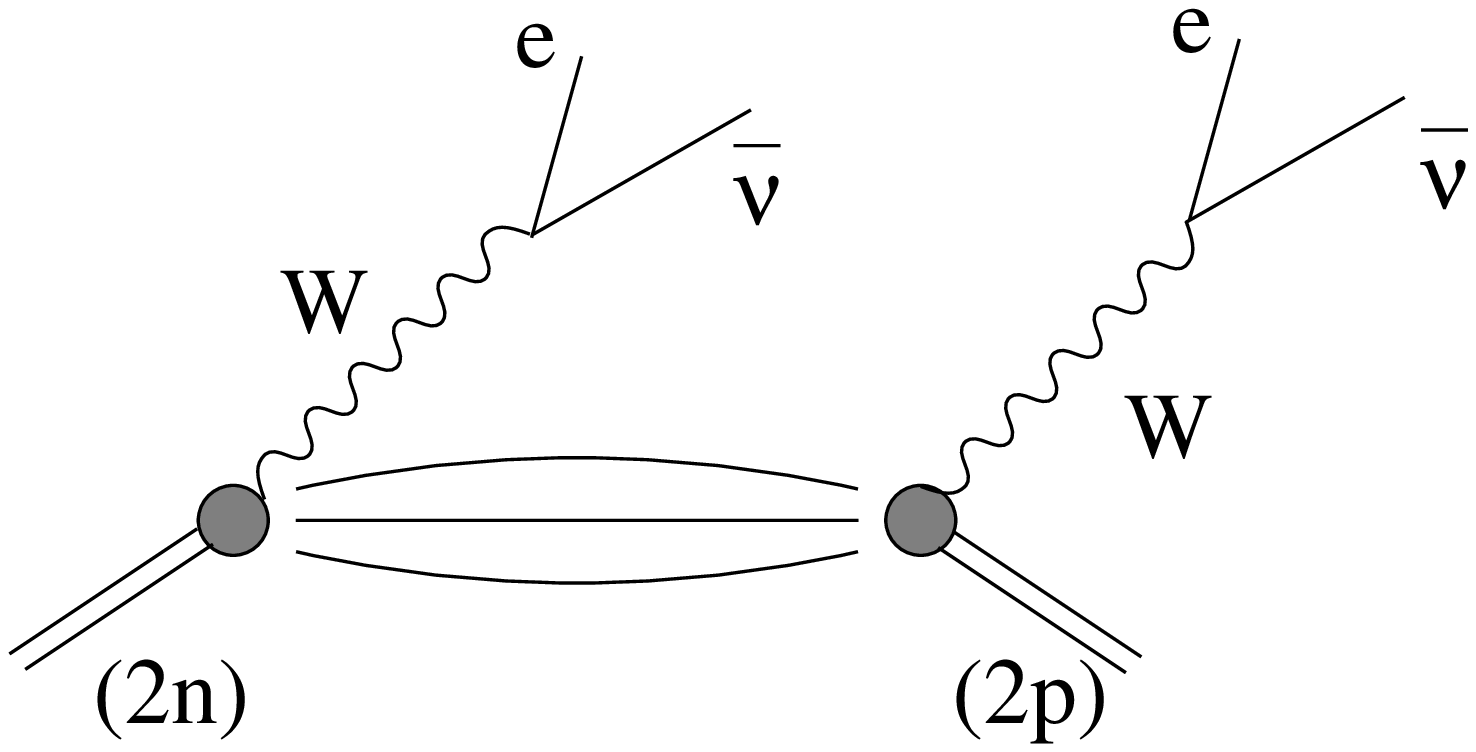}
}
\caption{}
\label{fig5}
\end{figure}

Direct counter experiments have observed $(2\beta )_{2\nu }$ in a variety of
nuclei, with half-lifes of the order $10^{19}-10^{24}$ years.

\section{Neutrino Oscillations}

The most sensitive method to prove that neutrinos are massive is provided by
neutrino oscillations $\left[ 20\right] $. Neutrino oscillations are quantum
mechanical processes based on mass and mixing of the neutrino flavours. If
the weak interaction states (greek indices) do not coincide with the mass
eigenstates (latin indices), the first ones are coherent superpositions of
definite mass states

\begin{equation}
\nu _{\alpha }=\sum_{k}U_{\alpha k}\nu _{k},  \label{eq:num36}
\end{equation}
where $\nu _{k}$ can be either Dirac or Majorana particles. For the present
discussion, we will limit ourselves to active left-handed $\nu _{k}$'s,
although the superposition (36) has to be extended to light sterile
neutrinos if they exist in Nature.

The propagation of the state $\nu _{\alpha }$ in vacuum, if it was prepared
as such at $t=0$, is

\begin{equation}
|\nu _{\alpha }(t)\rangle =\sum_{\beta }|\nu _{\beta }\rangle \left(
\sum_{k}U_{\beta k}^{\ast }e^{-iE_{k}t}U_{\alpha k}^{{}}\right) .
\label{eq:num37}
\end{equation}

The transition probability that (37) be observed, at a distance $L\simeq t$,
as $\nu _{\beta }$ is given by

\begin{equation}
P(\nu _{\alpha }\rightarrow \nu _{\beta })=\delta _{\beta \alpha }+2{\rm Re}%
\sum_{k<j}U_{\beta k}^{\ast }U_{\alpha k}^{{}}U_{\beta j}^{{}}U_{\alpha
j}^{\ast }\left( e^{-i\Delta m_{kj}^{2}\frac{L}{p}}-1\right) ,
\label{eq:num38}
\end{equation}
where $p\simeq E$ is the neutrino momentum $(\simeq energy)$ and $\Delta
m_{kj}^{2}$ the square mass differences of the physical neutrinos.

One realizes that the conditions

\begin{equation}
\Delta m_{kj}^{2}\frac{L}{E}<\!<1,\quad \forall k\neq j,  \label{eq:num39}
\end{equation}
if satisfied for all neutrinos, lead immediately to $P(\nu _{\alpha
}\rightarrow \nu _{\beta })\simeq \delta _{\beta \alpha }$. We conclude
that, in order to observe neutrino oscillations, in addition to mixing one
needs at least one $\Delta \ m^{2}$ with $\Delta \ m^{2}\gtrsim \frac{E}{L}$%
. The ``canonical'' sensitivity of some natural and person-made neutrino
sources is given in the Table

\bigskip

\begin{tabular}{llllll}
& {\small Sun} & {\small Atmosph} & {\small Reactors} & {\small Meson
Factories} & {\small H.E. Accel} \\
$\frac{E}{L}(eV)^{2}$ & {\small 10}$^{-11}$ & {\small 10}$^{-3}$ & {\small 10%
}$^{-2}$ & {\small 10}$^{-1}$ & {\small 1}
\end{tabular}

\bigskip

These values can however be modified by either long-base-line experiments
with intense neutrino beams or the effect of neutrino interactions in matter
(see next Section).

The flavour detection at distance L by means of charged current interactions
allows the classification of neutrino oscillation experiments into two types:

i) Appearance Experiments, described in Fig.6

\bigskip

\begin{figure}[h!]
\centerline{
\epsfxsize=6cm
\epsffile{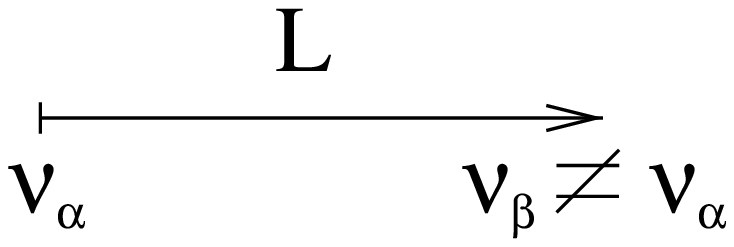}
}
\caption{}
\label{fig6}
\end{figure}

ii) Disappearance Experiments, in which one measures the Survival
Probability, as shown in Fig.7

\bigskip

\begin{figure}[h!]
\centerline{
\epsfxsize=6cm
\epsffile{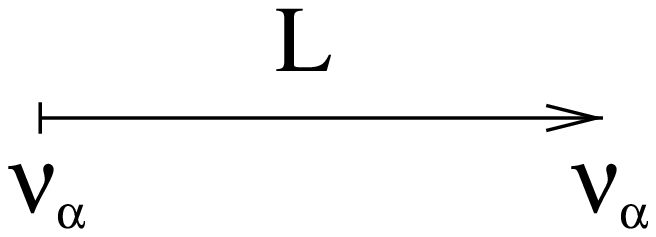}
}
\caption{}
\label{fig7}
\end{figure}

For low energy $\overline{\nu }_{e}(\nu _{e})$ neutrinos such as produced by
reactors (the sun), one is automatically limited to disappearance
experiments. Pure neutral current neutrino detection does not discriminate
among flavours, so that it is insensitive to neutrino oscillations. The
neutrino detection by elastic scattering on electrons is, in general,
dominated by charged current interactions with some proportion of neutral
current scattering. An interesting exception to this rule happens for
reactor antineutrinos $\overline{\nu }_{e}$ at energies around 0.5 $MeV$,
where a dynamical zero $\left[ 21\right] $ operates for $(\overline{\nu }_{e}
$ $e)$ scattering only, but not for $(\overline{\nu }_{\mu }$ $e)$ or $(%
\overline{\nu }_{\tau }e)$ scattering. The neutrino oscillation is then
manifested $\left[ 22\right] $ like an appearance experiment. By varying the
energy of the $\overline{\nu }_{e}$, one could tune the proportion of
appearance versus disappearance behaviours.

For an oscillation between two neutrino types, the mixing matrix of Eq.(36)
is real and orthogonal

\begin{equation}
U=\left(
\begin{array}{cc}
\cos \theta & \sin \theta \\
-\sin \theta & \cos \theta
\end{array}
\right) ,  \label{eq:num40}
\end{equation}
so that the appearance and survival probabilities are given,
correspondingly, by $(\nu ^{\prime }\neq \nu )$

\begin{equation}
\left.
\begin{array}{l}
P (\nu \to \nu^{\prime})=\frac{1}{2}\sin^2 2\theta \left ( 1-\cos \frac{%
\Delta m^2 L}{2 E} \right ) \\
P(\nu \to \nu)=1-P(\nu \to \nu^{\prime})
\end{array}
\right \} .  \label{eq:num41}
\end{equation}

\bigskip

With two intervening parameters $(\Delta m^{2},\sin ^{2}2\theta )$, the
analysis of neutrino oscillation experiments is presented in ``exclusion''
plots.

The general mixing for three families contains three angles and one CP
phase, accompanying two independent mass differences. All these ingredients
have to have an active participation in order to generate CP violating
observables $\left[ 23\right] $. A program of this kind needs intense beams
of high energy neutrinos with different flavours and well known spectra.

For a hierarchical spectrum of neutrinos $m_{1}<<m_{2}<<m_{3}$, we can
assume $\Delta m_{12}^{2}\frac{L}{2E}<<1$, except for solar or cosmic
neutrinos. Under the assumption of a single relevant $\Delta
m_{23}^{2}\simeq \Delta m_{13}^{2}$mass difference, the appearance
probability in the three-family case $(\beta \neq \alpha )$ becomes

\bigskip

\begin{equation}
P_{\alpha \rightarrow \beta }=2|U_{\beta 3}|^{2}|U_{\alpha 3}|^{2}\left(
1-\cos \frac{\Delta m_{23}^{2}L}{2E}\right) ,  \label{eq:num42}
\end{equation}
which has the same oscillating form as for the case of two neutrino types.
The effective ``mixing'' has however a different meaning. The survival
probability is given by

\bigskip

\begin{equation}
P_{\alpha\to\alpha}=1-\sum_{\beta\neq\alpha}P_{\alpha\to\beta}=
1-2|U_{\alpha 3}|^2(1-|U_{\alpha 3}|^2) \left (1-\cos\frac{\Delta m_{23}^2 L%
}{2 E} \right ).  \label{eq:num43}
\end{equation}

\bigskip

As a consequence, the Disappearance Reactor Experiments (CHOOZ, Palo Verde)
are primarily a measure of $\left| U_{e3}\right| $.

\section{Neutrino oscillations in matter}

In a medium, the electron-neutrinos $\nu _{e}$ acquire an extra inertia due
to the extra charged current interaction with the electrons of matter,
described by the first diagram of Fig.8

\begin{figure}[h!]
\centerline{
\epsfxsize=4cm
\epsffile{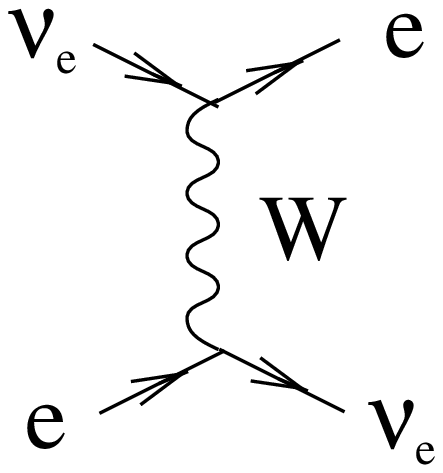}
\hspace{2cm}
\epsfxsize=5cm
\epsffile{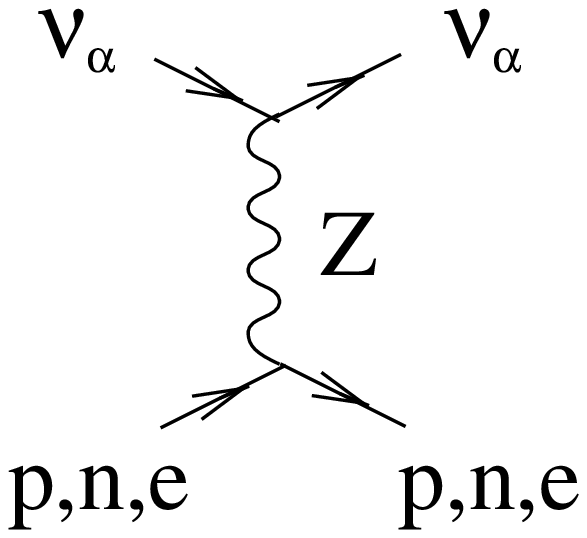}
}
\caption{}
\label{fig8-9}
\end{figure}

As a consecuence, in their propagation the $\nu _{e}$ acquires an extra
phase coming from $<H_{cc}>$. The vector charge density for electrons
contributes coherently to forward scattering, so that $<\overline{e}$ $%
\gamma ^{0}$ $e>=<e^{+}$ $e>=n_{e}$, with $n_{e}$ the electron number
density. The other terms of $H_{cc}$are not coherent, so that

\bigskip

\begin{equation}
\langle H_{cc}\rangle \approx \sqrt{2}G_{F}n_{e}  \label{eq:num44}
\end{equation}
for $\nu _{e}$. All flavours $(\nu _{e},\nu _{\mu },\nu _{\tau })$ have a
common neutral current interaction, described by the second diagram of
Fig.8, which leads to a common phase in their propagation. This common phase
is irrelevant.

In the flavour basis, the matrix elements of the Hamiltonian in vacuum are
given by $U_{v}$ $H_{v}$ $U_{v}^{+}$, where $(H_{v})_{ij}=(p+\frac{m_{i}^{2}%
}{2E})\delta _{ij}$ and $U_{v}$ is the mixing matrix. For two families, it
is given by Eq.(40) and the effective evolution equation, up to terms
proportional to identity, is

\bigskip 

\begin{equation}
i\frac{d}{dt} \left \{
\begin{array}{c}
\nu_e \\
\nu_{\mu}
\end{array}
\right \}= \left (
\begin{array}{cc}
-\frac{\Delta m^2}{4 E}\cos 2 \theta_v & \frac{\Delta m^2}{4 E}\sin 2
\theta_v \\
\frac{\Delta m^2}{4 E}\sin 2 \theta_v & \frac{\Delta m^2}{4 E}\cos 2 \theta_v
\end{array}
\right ) \left \{
\begin{array}{c}
\nu_e \\
\nu_{\mu}
\end{array}
\right \} .  \label{eq:num45}
\end{equation}

In matter, the effective hamiltonian in the flavour basis is obtained from
Eq. (45) by the addition of the diagonal non-universal charged current
matrix element (44). Once again, up to terms proportional to the identity
(as it is the case for the neutral current interaction), the evolution in
matter is governed by

\bigskip 

\begin{equation}
i\frac{d}{dt} \left \{
\begin{array}{c}
\nu_e \\
\nu_{\mu}
\end{array}
\right \}= \left (
\begin{array}{cc}
-\frac{\Delta m^2}{4 E}\cos 2 \theta_v +\sqrt{2}G_F n_e & \frac{\Delta m^2}{%
4 E}\sin 2 \theta_v \\
\frac{\Delta m^2}{4 E}\sin 2 \theta_v & \frac{\Delta m^2}{4 E}\cos 2 \theta_v
\end{array}
\right ) \left \{
\begin{array}{c}
\nu_e \\
\nu_{\mu}
\end{array}
\right \} .  \label{eq:num46}
\end{equation}

The diagonalization of the effective hamiltonian of Eq.(46) by means of

\bigskip 

\begin{equation}
U_{M}=\left(
\begin{array}{cc}
\cos \theta _{M} & \sin \theta _{M} \\
-\sin \theta _{M} & \cos \theta _{M}
\end{array}
\right)  \label{eq:num47}
\end{equation}
leads to the following solution for $\theta _{M}$, at constant density $%
n_{e} $,

\bigskip

\begin{equation}
\tan 2\theta _{M}=\frac{\tan 2\theta _{v}}{1-\frac{L_{v}}{L_{e}}\frac{1}{%
\cos 2\theta _{v}}},  \label{eq:num48}
\end{equation}
where $L_{e}$ is the interaction length of $\nu _{e}$ and $L_{v}$ is the
oscillation length in vacuum:

\bigskip

\begin{equation}
L_e=\frac{\sqrt{2} \pi}{G_F n_e}; \quad L_v=\frac{4 \pi E}{\Delta m^2}.
\label{eq:num49}
\end{equation}

The matter stationary eigenstates do not coincide with the mass eigenstates
in vacuum. It is remarkable from Eq.(48) that, independent of how small $%
\theta _{v}$ could be, $\theta _{M}$ can give maximal mixing $(=\frac{\pi }{4%
})$ if a ``resonance condition'' is satisfied as given by

\bigskip

\begin{equation}
\left (\frac{L_v}{L_e} \right )_{{\rm res}}=\cos 2\theta_v .
\label{eq:num50}
\end{equation}

This resonance enhancement constitutes the cellebrated MSW effect $\left[ 24%
\right] $. In view of Eqs.(49), the resonance condition appears at a
``resonance energy'' for $n_{e}$ and $\Delta m^{2}$ fixed. For the resonance
enhancement to be possible, $(\Delta m^{2}\cos 2$ $\theta _{v})$ has to be
positive for neutrinos. With the natural convention (1,2) for the order of
the mass eigenvalues, $\Delta m^{2}>0$, so that Eq.(50) needs

\bigskip

\begin{equation}
\cos 2\theta _{v}=\cos ^{2}\theta _{v}-\sin ^{2}\theta _{v}>0,
\label{eq:num51}
\end{equation}
i.e., the lowest mass eigenstate has to have a larger $\nu _{e}$ component.
In the case of antineutrinos, the coherent interaction amplitude (44)
changes sign, so that the resonance condition is only possible if $\cos 2$ $%
\theta _{v}<0$. We conclude that either neutrinos or antineutrinos, but not
both, can show the resonance enhancement.

The appearance probability for neutrinos in matter is dictated by the matter
mixing angle $\theta _{M}$ and the energy gap $E_{2}^{M}-E_{1}^{M}$ in
matter. One obtains

\bigskip

\begin{equation}
|\langle \nu _{\mu }|\nu _{e}(t)\rangle |^{2}=\sin ^{2}2\theta _{M}\sin ^{2}%
\frac{\pi L}{L_{M}(E)},  \label{eq:num52}
\end{equation}
where $L_{M}(E)$ is the energy dependent ``matter oscillation length''

\bigskip

\begin{equation}
L_M(E)=\frac{L_v}{\left [1-2\frac{L_v}{L_e}\cos 2\theta_v+ \left (\frac{L_v}{%
L_e} \right )^2 \right ]^{1/2}}.  \label{eq:num53}
\end{equation}

Two comments in connection with the oscillation probability (52):

( i ) The probability of mixing

\bigskip

\begin{equation}
\sin ^{2}2\theta _{M}=\frac{\left( \frac{\sin 2\theta _{v}}{L_{v}}\right)
^{2}}{\left( \frac{\cos 2\theta _{v}}{L_{v}}-\frac{1}{L_{e}}\right)
^{2}+\left( \frac{\sin 2\theta _{v}}{L_{v}}\right) ^{2}},  \label{eq:num54}
\end{equation}
has a typical resonance form, with the maximun $(=1)$ at the resonance
energy (50).

( ii) The oscillation phase does not depend explicitly on $\frac{L}{E}$
anymore, but the result (52) is still an even function of $L$.

When the matter has a varying density, like the case of neutrinos
propagating from the center of the sun, the evolution equation(46) cannot be
solved analitically, except for a few selected functional dependences. There
is an important case, however, in which one can discuss a simple approximate
solution: that of an adiabatically, slowly, varying density. Let us consider
electron neutrinos generated in a region of high density, like the center of
the sun, much higher that the value associated with the resonance condition.
Eq.(48) tells us that the mixing angle in matter is $\theta _{M}\simeq \frac{%
\pi }{2}$, i.e., $\nu _{e}$ is near the higher energy state in matter and
neutrino mixing is suppressed. As neutrinos propagate towards regions of
smaller density, $\theta _{M}$ decreases and eventually reaches the
resonance condition, in which neutrino mixing is maximal $\theta _{M}=\frac{%
\pi }{4}$. Further propagation to smaller densities, like for neutrinos
leaving the sun, leads to level crossing and $\theta _{M}\rightarrow \theta
_{v}$. If this is small, $\nu _{e}$ is near the lowest energy state and
neutrino mixing is again small. For adiabatic evolution, the system remains
in the corresponding stationary state in level ordering. If originally the
electron neutrino was in the upper level, the system will evolve to the
upper level of the modified hamiltonian, which at the end is near the muon
neutrino. We conclude that there is complete Flavour Conversion in this
particular case. This discussion is illustrated in  Fig.9, which shows the
energy levels as a function of the matter density

\begin{figure}[h!]
\centerline{
\epsfxsize=8cm
\epsffile{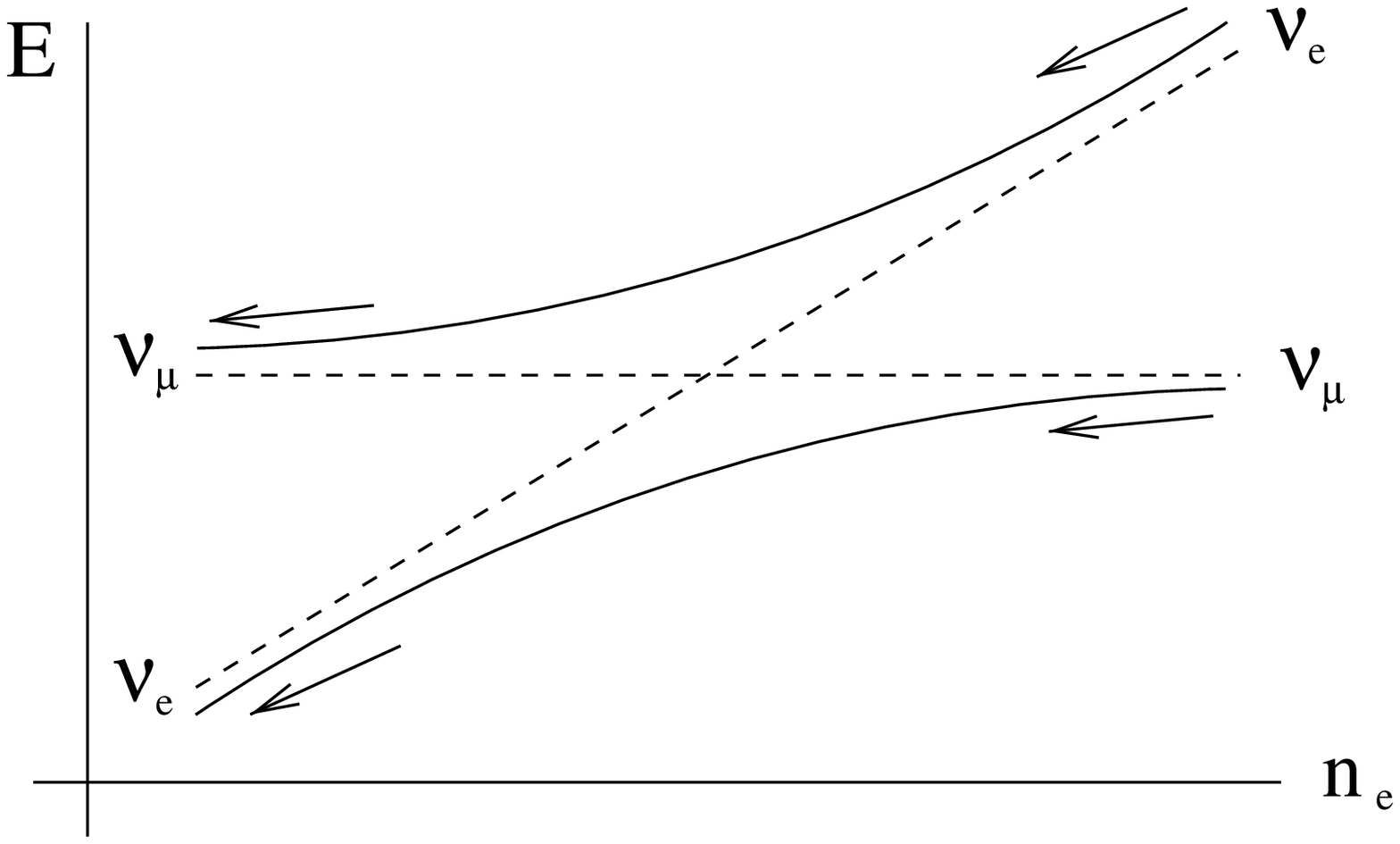}
}
\caption{}
\label{fig10}
\end{figure}

\section{Outlook}

The problem of neutrino masses and mixing has entered a very fascinating
era. More than fourty experiments are devoted to the field and the prospects
for positive results and definite answers look accessible.

Massive neutrinos can open a unique window to a new scale in physics. We
still do not know whether neutrinos are Dirac or Majorana particles.
Similarities and differences with the quark mixing can provide clues to the
flavour problem.

There are several subjects related to neutrino physics not covered in these
lectures. Among them, one can cite CP, and T, violation in the leptonic
sector, the connection to flavour changing neutral current processes,
electric and magnetic dipole moments, high energy neutrino astronomy. Some
of these topics can be followed in the excellent lectures by Akhmedov$\left[
25\right] $. The experimental status of neutrino oscillation studies is
discussed by Di Lella in these Proceedings.

\section*{Acknowledgments}

I am grateful to the organizers of the meeting for a very pleasant
environment and to Mari Carmen Ba\~{n}uls for a critical reading of the
manuscript. This work was supported by CICYT, under Grant AEN99-0962

\section*{References}

$\left[ 1\right] $ Y. Fukuda et al., Phys. Rev. Lett. 81 (1998) 1562; Phys.
Rev. Lett. 82(1999)2644; Phys. Rev. Lett. 82(1999)5194.

\noindent
$\left[ 2\right] $ E. Fermi, Nuovo Cimento 11(1934)1; Z. Phys. 88(1934)161

\noindent
$\left[ 3\right] $ F. Reines, C. Cowan, Phys. Rev. 113(1959)273

\noindent
$\left[ 4\right] $ L. Landau, Nucl. Phys. 3(1957)127; A. Salam, Nuovo
Cimento 5 (1957) 299; T.D. Lee, C.N. Yang, Phys. Rev. 105(1957)1671.

\noindent
$\left[ 5\right] $ M. Goldhaber, L. Grodzins, A.W. Sunyar, Phys. Rev. 109
(1958)1015

\noindent
$\left[ 6\right] $ E.C.G. Sudarshan, R.E. Marshak, Phys. Rev. 109(1958)1860;
R.P. Feynman, M. Gell-Mann, Phys. Rev. 109(1958)193; J.J. Sakurai, Nuovo
Cimento 7(1958)649

\noindent
$\left[ 7\right] $ Review of particle physics, C. Caso et al., Eur. Phys. J.
3(1998)1.

\noindent
$\left[ 8\right] $ See, for example, J. Bernab\'{e}u, P. Pascual,
``Electro-Weak Theory'', GIFT book, Univ. Aut. Barcelona, B-21399(1981)

\noindent
$\left[ 9\right] $ Z. Maki, M. Nakagava, S. Sakata, Progr. Theor. Phys.
28(1962)870.

\noindent
$\left[ 10\right] $ E. Majorana, Nuovo Cimento 14(1937) 171

\noindent
$\left[ 11\right] $ J. Bernab\'{e}u, P. Pascual, Nucl. Phys. B228(1983)21.

\noindent
$\left[ 12\right] $ S. Weinberg, Phys. Rev. Lett. 43(1979)1566.

\noindent
$\left[ 13\right] $ E. Ma, hep-ph 0001186; S.M.Barr, 003101.

\noindent
$\left[ 14\right] $T. Yanagida, in Proc. Workshop on Unified Theory and
Baryon number in the Universe, Eds. O. Sawada, A. Sugamoto, KEK, Tsukuba,
Japan (1979)95; M. Gell-Mann, P. Ramond, R. Slansky, in Supergravity, Eds.
P. Nieuwenhuizen, D.Z. Freedman (North Holland 1979) 315.

\noindent
$\left[ 15\right] $ F. Fogli et al., hep-ph/9912231; M.C.
Gonz\'{a}lez-Garc\'{i}a, C. Pe\~{n}a-Garay, hep-ph/0001129.

\noindent
$\left[ 16\right] $ M. Apollonio et al., Phys. Lett. B466(1999)415.

\noindent
$\left[ 17\right] $ W.H. Furry, Phys. Rev. 56(1939)1184.

\noindent
$\left[ 18\right] $ S.M. Bilenky, C. Giunti, W. Grimus, B. Kayser, S.T.
Petcov, Phys. Lett. B465(1999) 193; V. Barger, K. Whisnant, Phys. Lett.
B456(1999)194; F. Vissani, JHEP 9906(1999)022; H.V. Klapdor-Kleingrothaus,
H. P\"{a}s, A. Yu. Smirnov, hep-ph/0003219.

\noindent
$\left[ 19\right] $ L. Baudis et al. Phys. Rev. Lett. 83(1999)41.

\noindent
$\left[ 20\right] $ B. Pontecorvo, J. Exp. Theor. Phys. 33(1957)549

\noindent
$\left[ 21\right] $ J. Segura, J. Bernab\'{e}u, F.J. Botella, J.
Pe\~{n}arrocha, Phys. Rev. D49(1994)1633.

\noindent
$\left[ 22\right] $ J. Segura, J. Bernab\'{e}u, F.J. Botella, J.A.
Pe\~{n}arrocha, Phys. Lett. B335(1994)93.

\noindent
$\left[ 23\right] $ J. Bernab\'{e}u, in Weak Interaction and Neutrinos,
WIN'99, Eds. C.A. Dom\'{i}nguez and R.D. Viollier, p.227; A. De R\'{u}jula,
M.B. Gavela, P. Hern\'{a}ndez, Nucl. Phuys. B547(1999)21; K. Dick, M.Freund,
M. Lindner, A. Romanino, Nucl. Phys. B562(1999)29; J. Bernab\'{e}u, M.C.
Ba\~{n}uls, Nucl. Phys. B(Proc. Suppl.)87(2000)315.

\noindent
$\left[ 24\right] $ L. Wolfenstein, Phys. Rev. D17(1978)2369; S.P. Mikheyev,
A. Yu. Smirnov, Sov. J.Nucl. Phys. 42(1985)913.

\noindent
$\left[ 25\right] $ E. Kh. Akhmedov, FISIST/1-2000/CFIF, hep-ph/0001264.

\end{document}